\documentclass[prb,twocolumn,showpacs,preprintnumbers,amsmath,amssymb,floatfix]{revtex4}
\usepackage{graphicx}
\usepackage{float}
\usepackage{trfsigns}
\usepackage{color}

\newcommand \be{\begin{eqnarray}}
\newcommand \ee{\end{eqnarray}}

\newcommand \ba{\begin{align}}
\newcommand \eea{\end{align}}

\begin{document}
      \csname @twocolumnfalse\endcsname
\title{One-dimensional electron fluid at high density}
\author{Vinod Ashokan$^1$\footnote{Corresponding author Email: vinod.ashokan@pu.ac.in}, N.\ D.\ Drummond$^2$, and K.\ N.\ Pathak$^1$}
\affiliation{$^1$Centre for Advanced Study in Physics, Panjab University, Chandigarh 160014, India}
\affiliation{$^2$Department of Physics, Lancaster University, Lancaster LA1 4YB, United Kingdom}
\begin{abstract}
We calculate the ground-state energy, pair correlation function, static structure factor, and momentum density of the one-dimensional electron fluid at high density using variational quantum Monte Carlo simulation. For an infinitely thin cylindrical wire the predicted correlation energy is found to fit nicely with a quadratic function of coupling parameter $r_s$. The extracted exponent $\alpha$ of the momentum density for $k\sim k_F$ is used to determine the Tomonaga-Luttinger parameter $K_\rho$ as a function of $r_s$ in the high-density regime for the first time. We find that the simulated static structure factor and pair correlation function for infinitely thin wires agree with our recent high-density theory [K.\ Morawetz \textit{et al.}, Phys.\ Rev.\ B \textbf{97}, 155147 (2018)].
\end{abstract}
\maketitle

\section{Introduction}

The quasi-one-dimensional uniform electron fluid has received much attention due to its interesting and theoretically challenging behavior and potential technological applications.\cite{Giuliani05,Giamarchi04} The advancement of material fabrication technology has permitted the realization of quasi-one-dimensional systems in carbon nanotubes,\cite{Saito98,Bockrath99,Ishii03,Shiraishi03} semiconducting nanowires,\cite{Schafer08,Huang01} confined cold atomic gases,\cite{Monien98,Recati03,Moritz05} edge states in quantum Hall liquids,\cite{Milliken96,Mandal01,Chang03} and conducting molecules.\cite{Nitzan03}

In this work we focus on high-density one-dimensional (1D) electron gases, which are strongly correlated systems at all densities. Gold nanowires are an example of such a system. Experimentally one can create quantum wires by self-assembly from gold vapor. The gold atoms arrange themselves naturally on a stepped silicon surface to form a linear atomic chain.\cite{Nagao06} The linear electron density for such a system is $n_{\rm 1D} = 1/(2r_s a_B^\star )$, where $n_{\rm 1D}=1.3\times10^7$cm$^{-1}$, $r_s$ is the coupling parameter, and the effective Bohr radius is $a_B^\star=[(\kappa_{\rm Si}+1)/2](m_0/m^\star) a_B$, where $a_B=0.529$ {\AA}, the silicon dielectric constant is $\kappa_{\rm Si}=11.5$, and $m^\star$ is the effective mass. The coupling constant is $r_s=0.52$ for $m^\star=0.45m_0$,\cite{Altmann01} and $r_s=0.7$ for $m^\star=0.60 m_0$.\cite{Altmann01,Losio01} Another realization of a high-density 1D electron gas can be achieved in zigzag carbon nanotubes placed on a SrTiO$_3$ substrate, which has a high dielectric constant.\cite{Javey02,Kim04,Fogler05b}
These approaches stand in contrast to the traditional way of obtaining a strongly correlated regime by lowering the electron density $n$ (or, equivalently, by increasing the coupling parameter $r_s$) of a two- or three-dimensional electron gas. The latter approach suffers from the problem of localization by disorder because of random charges on the substrate. The advantage of the former approach is that the Coulomb potential is strongly screened and the interaction among electrons is less affected by disorder.
Potential applications of 1D electron systems include future silicon nanowire junctionless field-effect transistor technology.\cite{Colinge10,Mirza17,Javey02} In such systems many-body effects including electron-electron interactions play an extremely important role in electronic transport.

It may be noted that the transverse confinement of the electrons modifies the effective electron-electron interaction potential and hence also plays a role in determining the properties of 1D electron systems. The effects of different models of confinement have been studied by comparing theoretical plasmon dispersions with experimental results.\cite{Moudgil10} It has been found that the harmonic model of confinement with the proper dielectric constant of the substrate and effective mass of the electron wire material provides the closest description of wires fabricated on substrates.

The effects of interactions in one-dimensional (1D) physics are radically different from those in higher-dimensional physics. The famous Landau conventional Fermi liquid theory\cite{Landau58,Nozieres61} is not applicable in 1D due to the fact that single-particle excitation energies and their inverse lifetimes are of the same order of magnitude. Further, the strength of these excitations is vanishingly small at low energies. Therefore the prospect of observing non-Fermi-liquid features has given a large impetus to both theoretical and experimental research on 1D materials. Electron-like quasiparticle excitations are distinctive attributes of higher-dimensional physics, whereas in 1D such individual excitations do not exist. The interaction in 1D turns the excitations into collective excitations, which are analogous to density oscillations (spin or electronic density). The theory describing the physical properties of 1D interacting systems (fermions, bosons, or spins) is the Tomonaga-Luttinger (TL) liquid.\cite{Tomonaga50,Luttinger63,Haldane81} The reduced dimensionality qualitatively changes the role of interactions, leading to phenomena such as spin-charge separation,\cite{Auslaender05} charge fractionalization,\cite{Steinberg08} and Wigner crystallization.\cite{Deshpande08}

The basic behavior of a TL liquid is characterized by key parameters known as the TL parameters or correlation exponents $K_\rho$ and $K_\sigma$. Together with the charge and spin collective excitation velocities $v_\rho$ and $v_\sigma$, these parameters completely determine the low-energy physics of the TL liquid.\cite{Voit97} In the absence of interactions $v_\rho$ and $v_\sigma$ reduce to the Fermi velocity, $v_F$, and the TL parameters are equal to unity. For fermions $K > 1$ corresponds to attractive interactions and $K < 1$ repulsive interactions.\cite{Schulz00} For spin-rotation invariant ($K_\sigma=1$) systems, all exponents can be expressed in terms of one parameter, $K_\rho$.  The spin-sector $K_\sigma\neq 1$ is possible only if spin rotation invariance is broken. The salient properties of a TL liquid are\cite{Voit97} (i) a continuous momentum distribution function $n(k)$ varying as $| k - k_ F |^\alpha$ around the Fermi wavenumber $k_F$, with $\alpha$ being the interaction-dependent exponent, and a pseudogap in the single-particle density of states $N(\omega)$ that is proportional to $|\omega|^{\alpha}$, which is a consequences of the absence of fermionic quasiparticles; (ii) correlation functions follow similar nonuniversal power laws, with interaction-dependent exponents; and (iii) spin-charge separation, with spin and charge degrees of freedom propagating at different velocities.

The exactly solvable TL liquid model describes electron correlations with short-range interactions. However, the electrons in a real 1D homogeneous electron gas (HEG) interact via a long-range Coulomb interaction. The long-range character of the Coulomb potential has been studied by Schulz.\cite{Schulz93} Further, the mapping of the long-range Coulomb interaction onto an exactly solvable model with short-range behavior has been studied by Fogler.\cite{Fogler05a,Fogler05b} In spite of these works the effects of long-range behavior within the TL model are still not fully understood.
The harmonically transversally confined wire with finite width $b$ has been studied with a lattice-regularized diffusion Monte Carlo technique by Casula \textit{et al.},\cite{Casula06} and by others.\cite{Shulenburger08,Malatesta00,Malatesta99} Several theoretical works have investigated the 1D HEG within Fermi liquid theory,\cite{Capponi00,Fano99,Poilblanc97,Valenzuela03,Fabrizio94,Friesen80,Calmels97,Garg08,Tas03,Renu12,Renu14,vinod18} and have been compared with the simulation results of Casula \textit{et al.}\cite{Casula06}\ with limited success.

Recently, the ground-state properties of the 1D electron liquid for an infinitely thin wire, and the harmonic wire have been studied using the quantum Monte Carlo (QMC) method by Lee and Drummond\cite{Lee11} for coupling parameters $r_s\geq1$.
%
%
%
%
In the present work we study the ground-state properties of infinitely thin and transversally confined harmonic wires using QMC as implemented in the \textsc{casino} code\cite{Needs10} in the high-density regime, i.e., $r_s<1$. The realistic long-range Coulomb interaction is taken to be $1/|x|$, as studied by Lee and Drummond.\cite{Lee11} From the simulated momentum density (MD) we have extracted its exponent $\alpha$ around $k\sim k_F$, which allows us to find the TL parameter $K_\rho$ as a function of $r_s$ in the high-density regime for the first time. The TL parameter $K_\rho$ we obtain at high density smoothly goes over to the value we obtained for low density from the data of Lee and Drummond.\cite{Lee11}

It is found that variational quantum Monte Carlo (VMC) correlation energies vary quadratically with $r_s$ in the high-density limit. Further, the simulated static structure factor (SSF) and pair correlation function (PCF) for infinitely thin wires are found to agree with our recent high-density theory.\cite{KM18}

The paper is organized as follows. The theoretical models used in our calculations for infinitely thin and harmonic wires are described in Sec.\ \ref{Theoreticalmodel}. In Sec.\ \ref{Computationalmethod} we outline our computational methodology, and provide the details of our approach. In Sec.\ \ref{Resultanddiscussion} we present results and discussion pertaining to the ground-state properties of the 1D HEG at high density. Finally, our overall conclusions are drawn in Sec.\ \ref{Conclusion}. In this article we use Hartree atomic units $(\hbar=|e|=m_e=4\pi\epsilon_0=1)$ throughout.


%

\section{Theoretical model}
\label{Theoreticalmodel}

The Hamiltonian for an $N$-electron 1D HEG is\cite{Lee11}
\begin{eqnarray}
\hat{H}=-\frac{1}{2}\sum_{i=1}^{N}\frac{\partial^2}{\partial x_i^2}+\sum_{i<j}V(x_{ij})+\frac{N}{2} V_{\rm Mad},
\end{eqnarray}
where $V(x_{ij})$ and $V_{\rm Mad}$ are the Ewald interaction and Madelung energy respectively. The Ewald interaction of an electron at $x_i$ with an electron at $x_j$ in an infinitely thin wire, determined by using Euler-Maclaurin summation,\cite{Saunders94} is
\begin{eqnarray}
V(x_{ij})=\sum_{n=-\infty}^{\infty}\bigg(\frac{1}{|x_{ij}+n L|} -\frac{1}{L}\int_{-L/2}^{L/2}\frac{dy}{|x_{ij}+n L -y|}\bigg),
\label{int_infinite_wire}
\end{eqnarray}
and for the harmonically confined wire the Ewald-like interaction potential is\cite{Lee11}
\begin{eqnarray}
V(x_{ij})&=&\sum^{\infty}_{m=-\infty}\bigg[ \frac{\pi}{2b} e^{(x_{ij}-mL)^2/(4b)^2} {\rm erfc}\bigg(\frac{|x_{ij}-mL}{2b}\bigg)\nonumber\\
& &-\frac{1}{|x_{ij}-mL|} {\rm erf}\bigg(\frac{|x_{ij}-mL}{2b} \bigg)\bigg]\nonumber\\
& &+\frac{2}{L}\sum^{\infty}_{n=1} {\rm E}_1[(bGn)^2]\cos(Gnx_{ij}),
\label{int_har_wire}
\end{eqnarray}
where $b$ is the width of the wire in units of the Bohr radius, $G=2\pi/L$, and ${\rm E}_1$ is the exponential integral function.
The Madelung constant $V_{\rm Mad}$ is the electrostatic potential at one electron due to all its periodic images (excluding itself); e.g., for the infinitely thin wire the Madelung constant is
\begin{eqnarray}
V_{\rm Mad}=\lim_{x\rightarrow 0}\bigg[ V(x)-\frac{1}{|x|}\bigg].
\end{eqnarray}

\section{Computational method}
\label{Computationalmethod}

We have computed the ground-state properties of the 1D HEG at high density $r_s<1$ using the VMC method. The details of the calculation method have been discussed by Lee and Drummond.\cite{Lee11}
A Slater-Jastrow-backflow trial wave function has been used in the calculation,\cite{Drummond04}
%
For a ferromagnetic system, the orbitals in the Slater determinants were plane waves with wavenumbers $k \lessapprox k_F=\pi/(2r_s)$. Using a backflow transformation\cite{Lopez06} provides an effective way to describe the three-body correlations in a 1D HEG\@. The free parameters in the trial wave function were optimized by unreweighted variance minimization,\cite{Umrigar88,Kent99,Drummond05} and energy minimization.\cite{Umrigar07}
The properties of the Slater-Jastrow-backflow trial wave function have been discussed in detail in Ref.\ \onlinecite{Lee11}, where it has been concluded that for infinitely thin wires the curvature of the wave function does not cancel the divergence in the interaction potential; hence the trial wave function possesses nodes at all of the coalescence points (i.e., at both like $\uparrow\uparrow$ and unlike $\uparrow\downarrow$ spin pairs). For infinitely thin 1D wires the ground-state energy is independent of the spin polarization, i.e., paramagnetic and ferromagnetic states are degenerate, which implies the inapplicability of the Lieb-Mattis theorem,\cite{Lieb62} whereas the ground-state energy is dependent on density. In the case of the harmonic wire, the paramagnetic and ferromagnetic sates are nondegenerate and there is nontrivial dependence on the spin polarization $\zeta=|N_\uparrow-N_\downarrow|/N$. However, in this work we restrict ourselves to calculations for the fully spin-polarized ($\zeta=1$) ferromagnetic fluid.

We have performed our calculations using only VMC\@. This is due to the fact that the VMC and diffusion Monte Carlo (DMC) results differ insignificantly. For example, for $r_s \geq 1$, Lee and Drummond\cite{Lee11} found VMC to retrieve more than 99.999\% of the correlation energy, while the VMC and extrapolated DMC estimates of the MD differ by no more than $\sim 2$ error bars.

\begin{figure}[!t]
\centering
\includegraphics[scale=0.3]{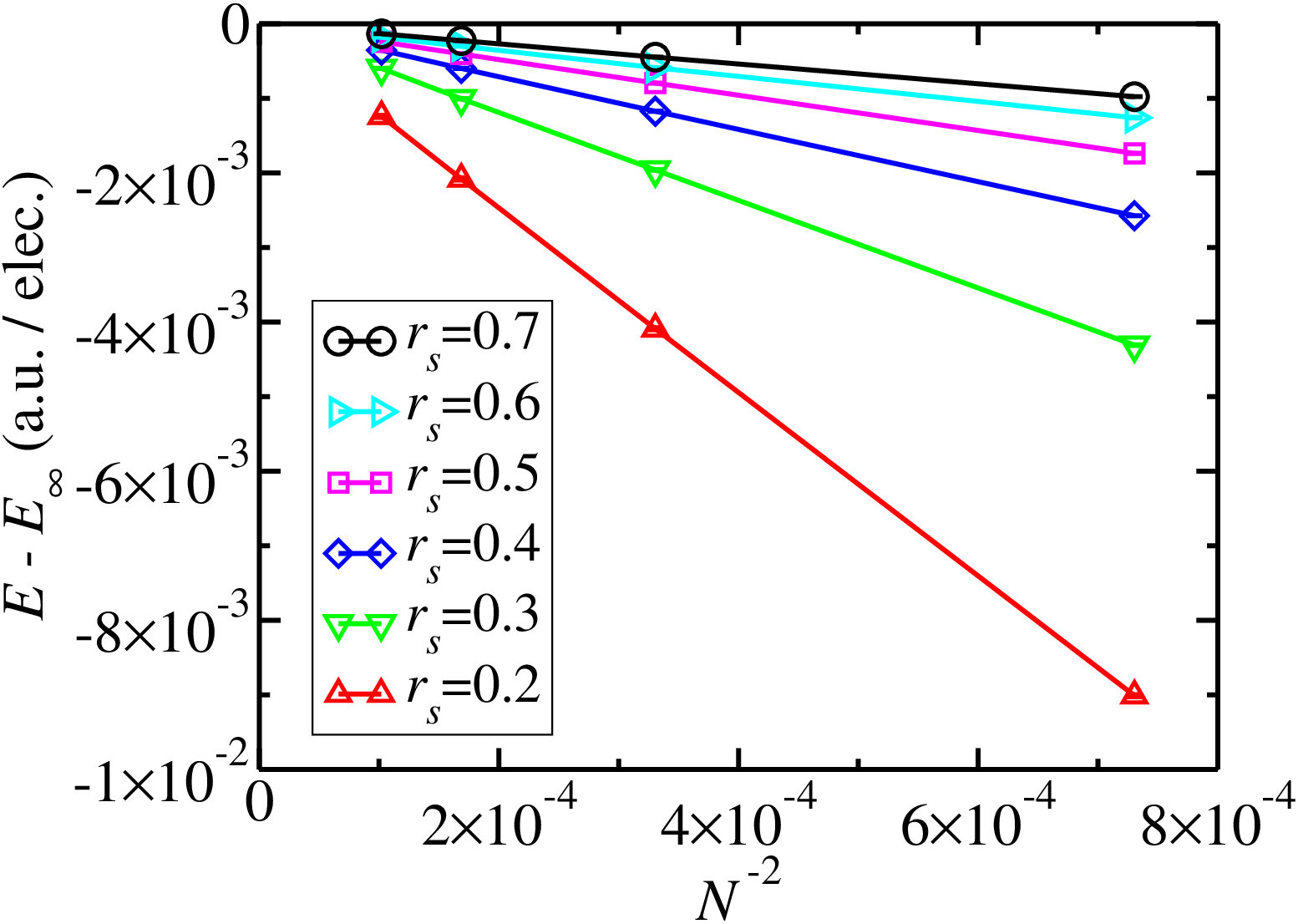}
\caption{(Color online) VMC energy $(E-E_{\infty})$ plotted against the reciprocal of the square of the system size for the infinitely thin wire. The energies per particle were extrapolated to the thermodynamic limit using the form $E(N)=E_{\infty}+BN^{-2}$.}\label{extrapolation}
\end{figure}

\section{Result and discussion}
\label{Resultanddiscussion}

{\bf Ground-state energy:} We have calculated VMC ground-state energies for $r_s= 0.1$, 0.2, 0.3, 0.4, 0.5, 0.6, 0.7, 0.8, and 0.9, with the number of electrons being $N=37$, 55, 77, and 99 for both infinitely thin and harmonic wires, and we have extrapolated to the thermodynamic limit. These results are given in Table \ref{gs_energy}. For infinitely thin wires the energies per particle were extrapolated to the thermodynamic limit using the form $E(N)=E_{\infty}+BN^{-2}$, where $B$ and $E_{\infty}$ are fitting parameters, which was derived in Ref.\ \onlinecite{Lee11} using the method proposed in Ref.\ \onlinecite{Chiesa06}. This form is also in agreement with conformal field theory results.\cite{Affleck86,Blote86} We used the same functional form for finite-size extrapolation of the energies of harmonic wires. Figure \ref{extrapolation} shows that this form fits the energy per particle of an infinitely thin wire well, and allows extrapolation to the thermodynamic limit. We used a Slater-Jastrow-backflow wave function for both infinitely thin and finite-thickness systems at high density, as used by Lee and Drummond\cite{Lee11} for low-density calculations. For the infinitely thin wire with $r_s=0.9$, $N=99$, and $\zeta=1$, the error bars on the ground-state energy were $O(10^{-7})$ a.u.\ per electron. The correlation energies for infinitely thin and harmonic wires are calculated from the VMC ground-state energy.

\begin{widetext}
The exchange energy is obtained\cite{vinod18} as
\begin{equation}
\epsilon_x(r_s,\zeta) = -\frac{1}{8r_s}\bigg\{ (1+\zeta)^2 \bigg[\frac{3}{2}-\gamma+\beta-\ln\bigg(\frac{\pi(1+\zeta)}{4r_s}\bigg) +\mathcal{L}\bigg]+
(1-\zeta)^2\bigg[\frac{3}{2}-\gamma+\beta-\ln\bigg(\frac{\pi(1-\zeta)}{4r_s}\bigg) +\mathcal{L}\bigg]\bigg\},
\label{ex_har_cyl}
\end{equation}
where $\zeta$ is the spin polarization, $\gamma$ is Euler's constant and $\mathcal{L}=-\ln(b)$. For infinitely thin wire $\beta=0$ and for thin harmonic wire $\beta=(\gamma/2)-\ln(2)$. Note that the logarithmic thickness of the wire is defined by $\mathcal{L}^{-1}$, and $\mathcal{L}$ cancels with the neutralizing background in the thermodynamic limit.
\end{widetext}


\begin{figure}[!t]
\centering
\includegraphics[scale=0.3]{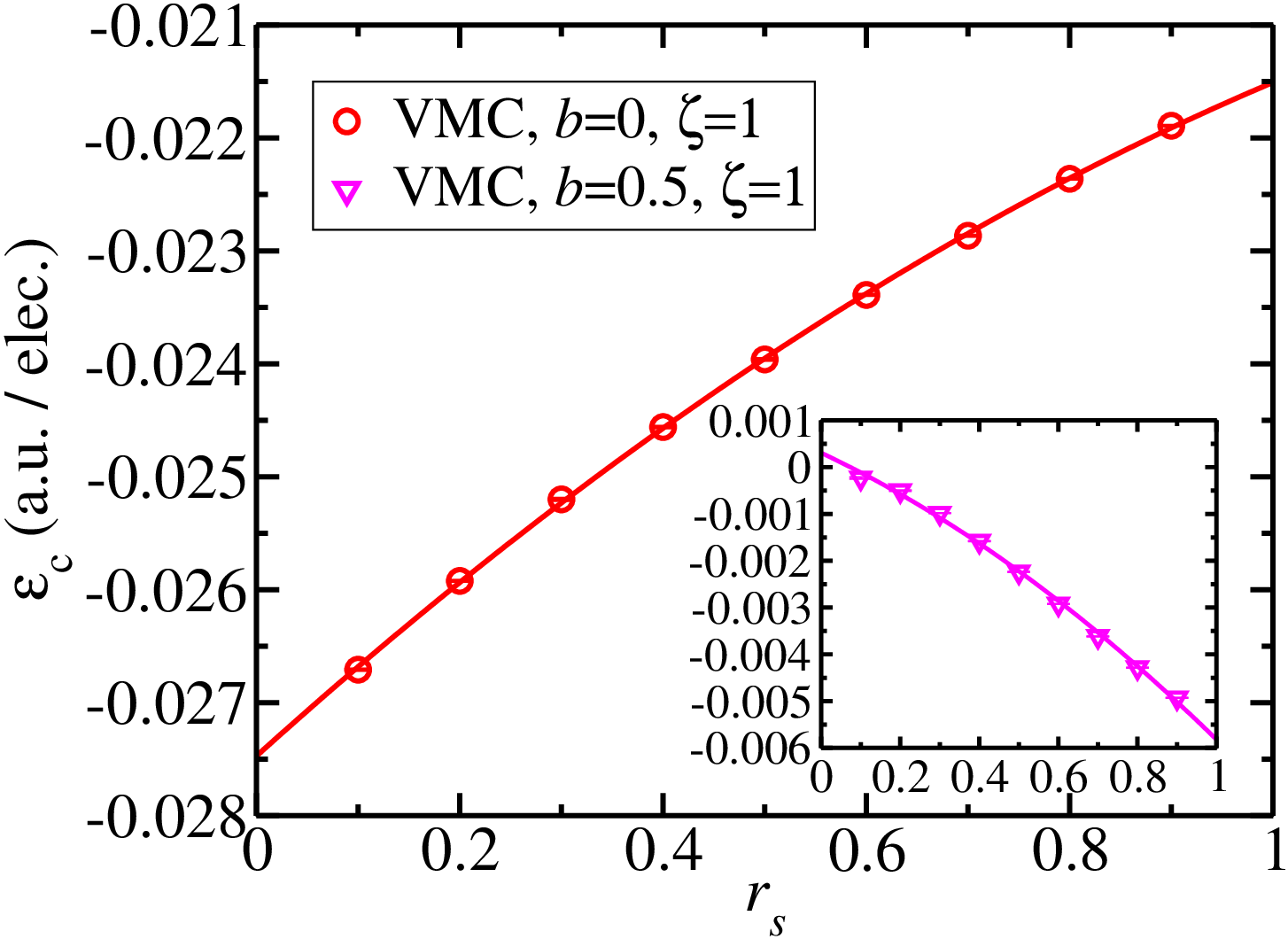}
\caption{(Color online) Correlation energy $\epsilon_c$ as a function of $r_s$ for an infinitely thin wire (and in the inset for a harmonic wire of width $b=0.5$). The solid line is a quadratic fit as a function of $r_s$.}\label{Corr_Energy}
\end{figure}

In Fig.\ \ref{Corr_Energy}, the correlation energy is plotted against $r_s$, and it is fitted with the quadratic function $\epsilon_c(r_s) = -0.027431(3) +0.00791(1)\; r_s-0.00196(1)\; r_s^2 $, where the constant and coefficient are in agreement with the conventional perturbation theory result,\cite{Loos13} i.e., $\epsilon_c(r_s)=-0.0274156+0.00845\;r_s+\ldots$ for high densities.
The correlation energy for the harmonic wire of width $b=0.5$ in the high-density limit is fitted with a quadratic function $\epsilon_c(r_s,b=0.5)=0.00079(1)-0.00581(4)\;r_s - 0.00062(3)\; r_s^2$.  This fit is accurate for $0.3\leq r_s <1$; at lower $r_s$, the fitted correlation energy (but not the raw data in Table \ref{gs_energy}) becomes unphysically positive.


\begin{table}
\begin{center}
\caption{\label{gs_energy} VMC ground-state total energies $E_\infty$ extrapolated to the thermodynamic limit and correlation energies $\epsilon_c$ for the infinitely thin wire ($b=0$) and harmonic wire ($b=0.5$).}
\begin{ruledtabular}
\begin{tabular}{ccccc}
&\multicolumn{2}{c}{Infinitely thin wire}&\multicolumn{2}{c}{Harmonic wire}\\
\cline{2-5}
&  $E_{\infty}$&$\epsilon_c$&$E_{\infty}$& $\epsilon_c $\\
\raisebox{1.5ex}[0pt]{$r_s$} & (a.u./elec.) & (a.u./elec.) & (a.u./elec.) & (a.u./elec.) \\
\hline
0.1& 50.25356(3) & $-0.026707(3)$   & 40.31883(2)      &   $-0.00023 (2) $ \\
0.2& 13.10051(1) & $ -0.025921(1)$   & 9.52997(1)      &   $-0.00050(1)$\\
0.3& 5.765347(7) & $-0.025199(7)$  & 3.861399(5)      &  $-0.000981(5)$  \\
0.4& 3.102011(5) & $-0.024559(5)$  & 1.898573(5)      &  $-0.001581(5)$ \\
0.5& 1.842920(5) & $-0.023960(5)$  & 1.004636(4)     &  $-0.002239(4)$  \\
0.6& 1.151946(3) & $-0.023389 (3)$  & 0.529594(3)      &   $-0.002924 (3)$  \\
0.7& 0.734582(3) & $-0.022865(3)$  & 0.251097(3)      &  $-0.003616 (3)$ \\
0.8& 0.465155(1) & $-0.022361(1)$  & 0.076506(2)      &  $-0.004285 (2)$  \\
0.9&0.2825597(9) & $-0.0218909(9)$ & $-0.038302(2)$  & $-0.004929(2)$ \\
\end{tabular}
\end{ruledtabular}
\end{center}
\end{table}

{\bf Pair correlation function:}
The parallel-spin PCF is defined as
\begin{eqnarray}
g_{\uparrow\uparrow}(x)=\frac{1}{\rho^2_{\uparrow}}\bigg<\sum^{N_{\uparrow}}_{i>j}\delta(|x_{i,\uparrow}-x_{j,\uparrow}|-x)\bigg>,
\label{pcf_up_spin}
\end{eqnarray}
where $\rho_{\sigma}$ is the electron density for spin $\sigma$, $x_{i,\uparrow}$ is the position of the $i^{th}$ electron with spin $\sigma$, and the angular brackets $\langle \cdots \rangle$ denote an average over configurations distributed as the square modulus of the wave function.

\begin{figure}[!t]
\centering
\includegraphics[scale=0.31]{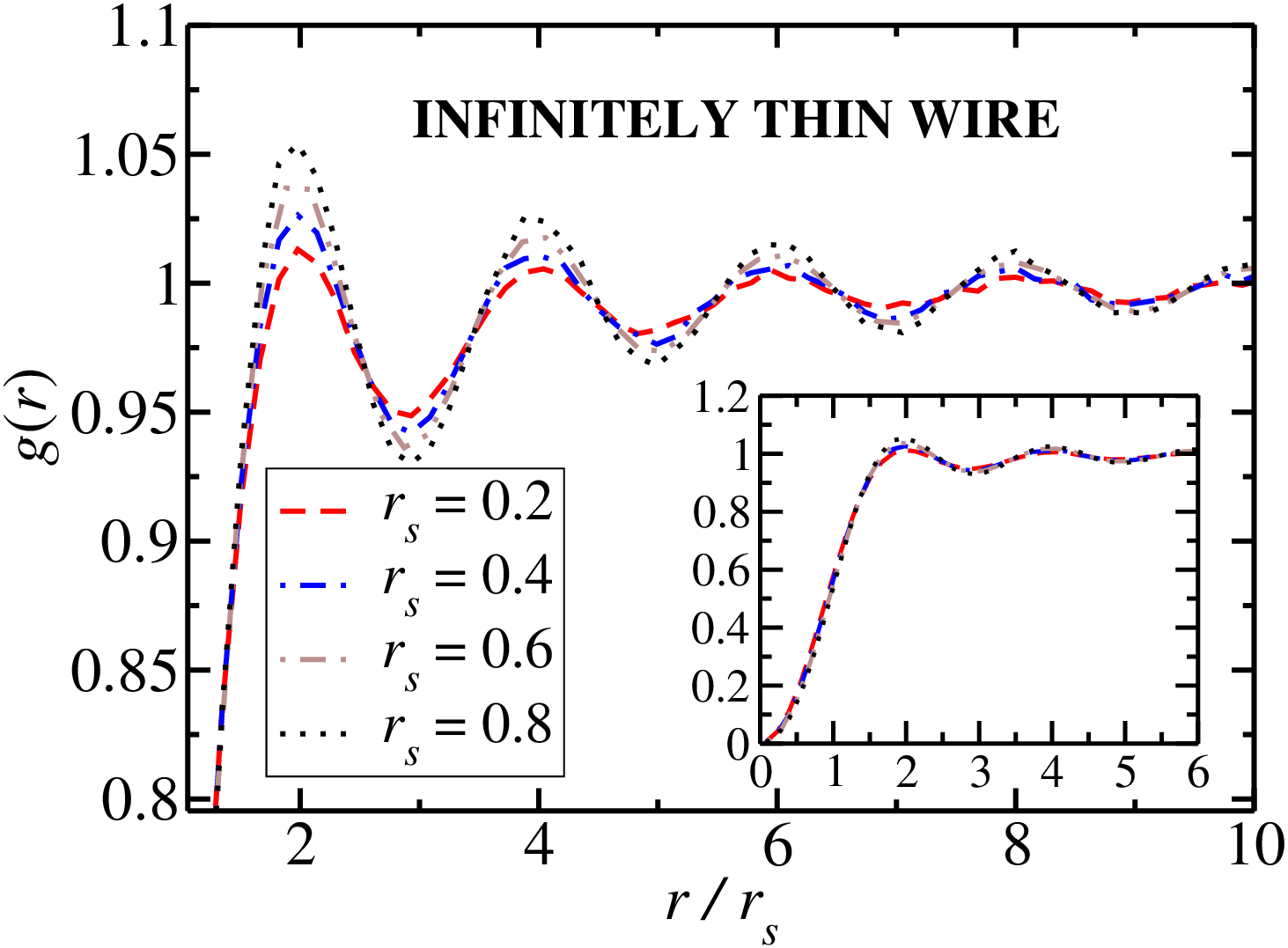}
\vskip 0.4in
\includegraphics[scale=0.31]{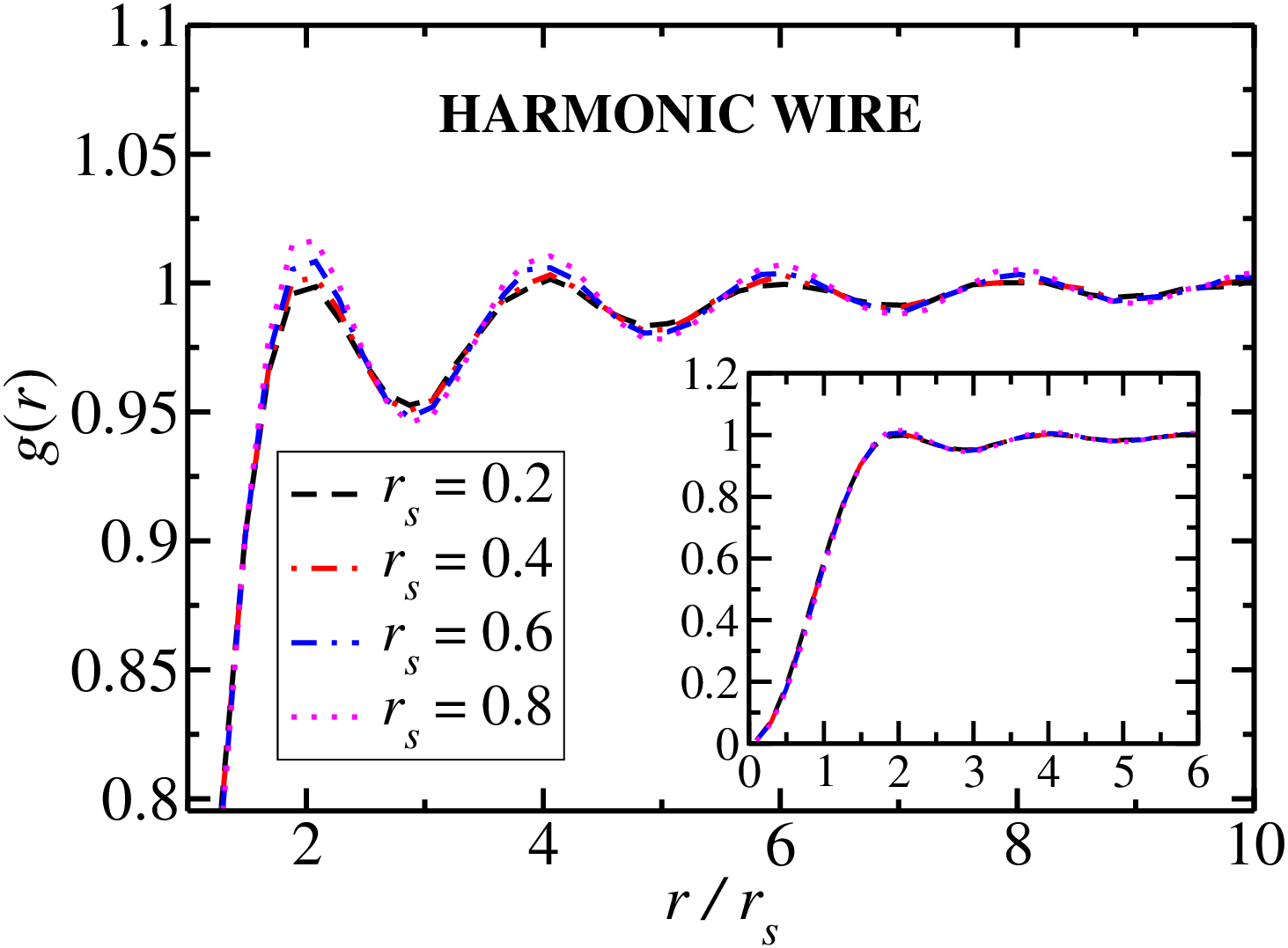}
\caption{(Color online) (upper) PCF of an infinitely thin wire at
several densities. (lower) PCF of a harmonic wire of width $b=0.5$ at
several densities. The data shown are for $N = 99$. The
inset shows the same data at the origin. The random errors in the data are small, as evidenced by the lack of visible noise in the curves.}\label{pcf_b0}
\end{figure}

\begin{figure}[!h]
\centering
\includegraphics[scale=0.33]{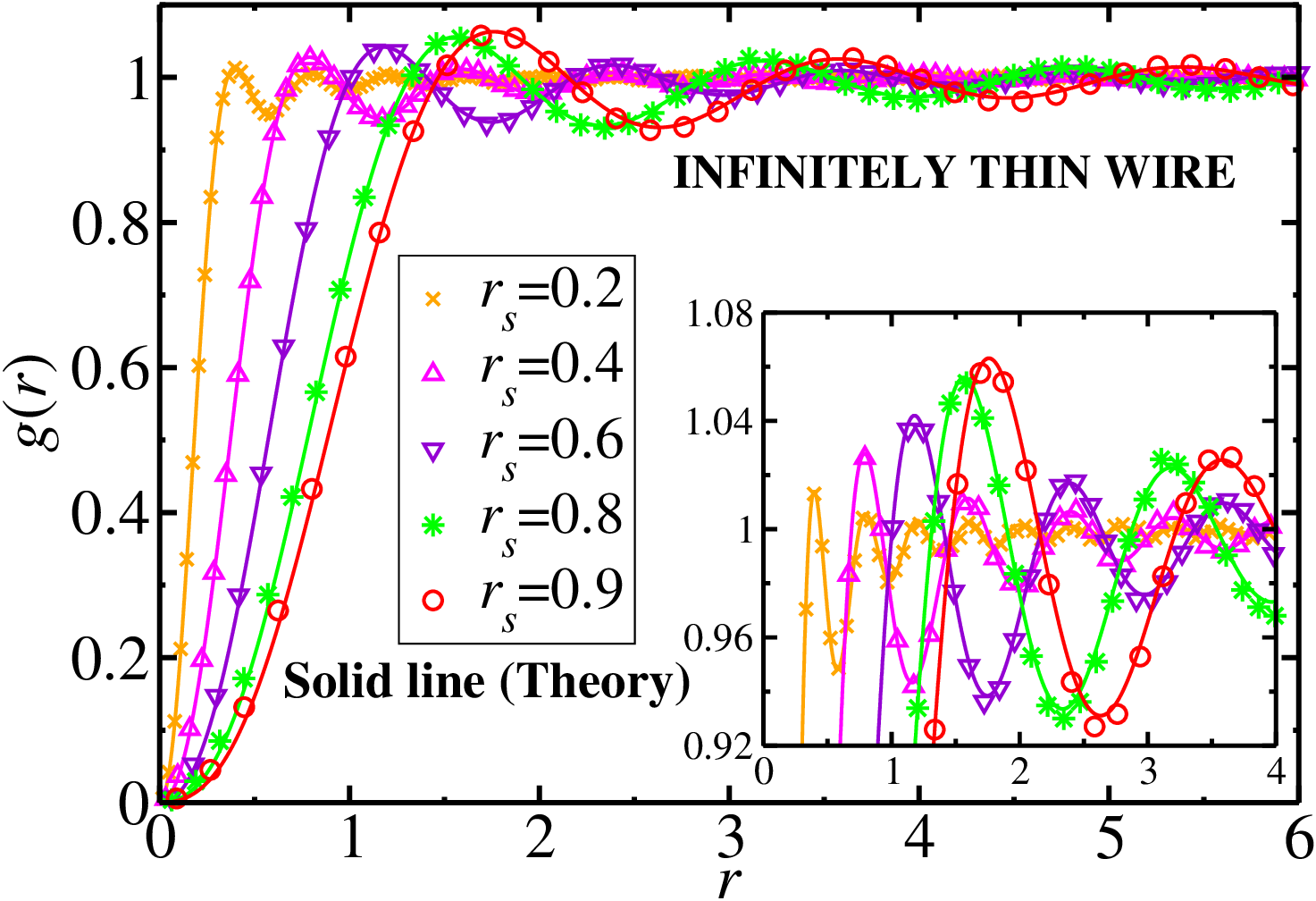}
\caption{(Color online) PCFs $g(r)$ of 1D HEGs in infinitely thin wires.  Results obtained from VMC simulation are compared with recent high-density theory\cite{KM18} at several densities. The inset shows the oscillations of $g(r)$ at larger distances in greater detail.}\label{gr_QMC_theory}
\end{figure}

%
The PCFs for an infinitely thin wire and a harmonic wire of thickness $b=0.5$ are plotted in Fig.\ \ref{pcf_b0} for several values of $r_s<1$. In Fig.\ \ref{gr_QMC_theory} we have compared the simulation with our recent high-density theory\cite{KM18} for infinitely thin wires, which was obtained in the $b\rightarrow 0$ limit for cylindrical wires. It may be recalled that the random phase approximation is known to lead to negative values of $g(r)$ at small distances,\cite{KM18} and this artifact is removed in the high-density theory which takes into account the vertex correction.

{\bf Static structure factor:} The SSF $S(k)$ is defined in terms of the PCF as
\begin{eqnarray}
S(k)=1+\frac{N}{L}\int[g(x)- 1]e^{-ikx} \, dx,
\label{ssf_k}
\end{eqnarray}
where $g(x)$ is $g_{\uparrow\uparrow}(x)$ in this case. The SSF contains information about the phase of the system. By using Eq.\ (\ref{pcf_up_spin}) in Eq.\ (\ref{ssf_k}) we obtain an expression for $S(k)$ which shows that it is a measure of the average squared amplitude of density fluctuations of wavenumber $k$.

\begin{figure}[!h]
\centering
\includegraphics[scale=0.35]{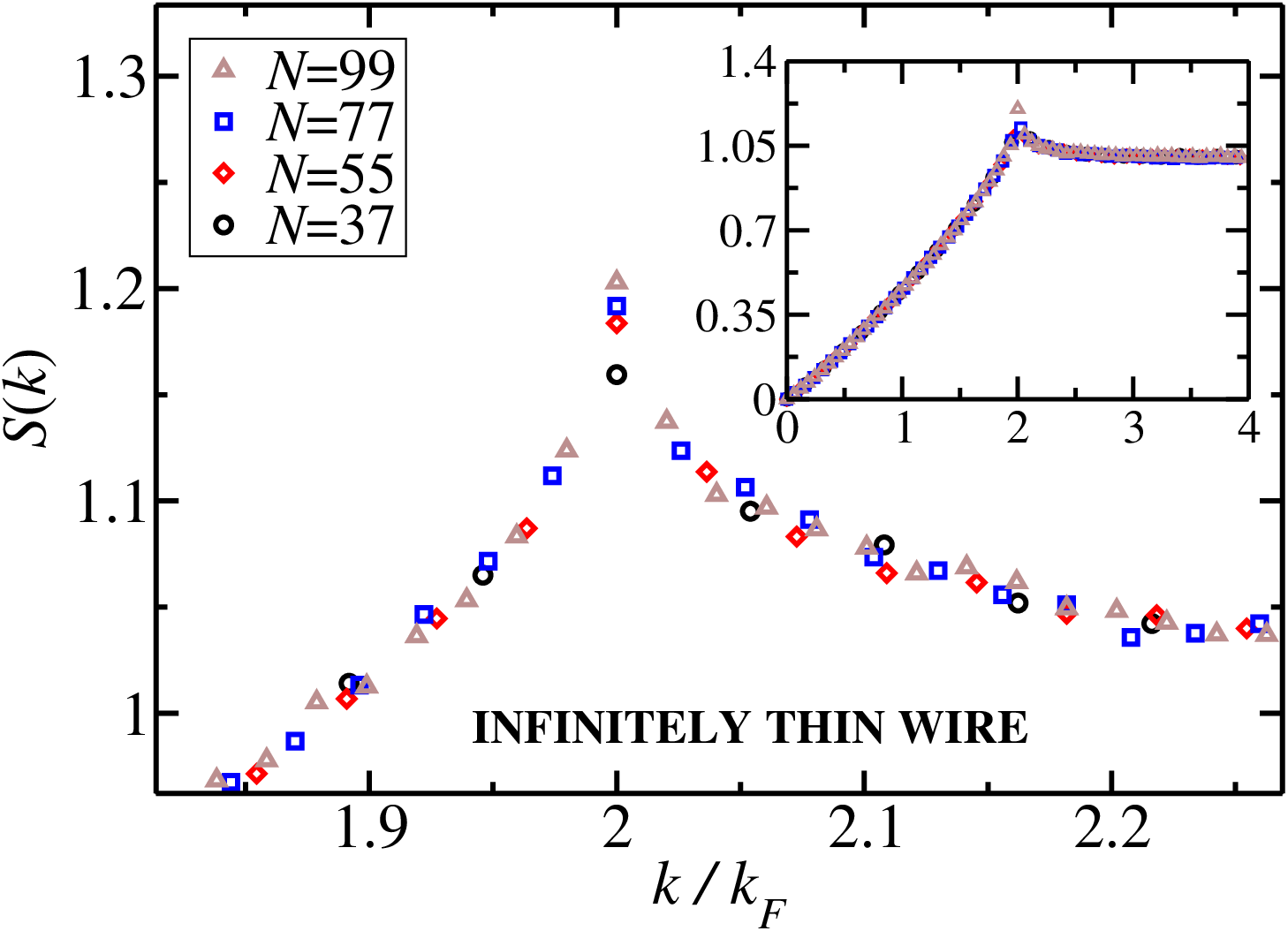}
\caption{(Color online) SSF of an infinitely thin wire at
 several system sizes for $r_s=0.8$. The main plot shows the behavior at the peak and the inset shows a zoomed-out view.}\label{ssf}
\end{figure}

The most interesting physics comes out of the behavior of the SSF at $k=2k_F$, because this corresponds to fluctuations with period $2r_s$, which
is the average interelectron spacing.  The height of the $2k_F$ peak in finite-cell SSFs does not scale as $N$ but appears to be sublinear,\cite{Lee11} as shown in Fig.\ \ref{ssf}, which is consistent with the presence of quasi-long-range order.  In this work we confirm that the SSF as a function of the wavenumber shows the typical uncorrelated behavior at high density with the double Fermi wavenumber $2k_F$ as characteristic inverse length. Also we have compared the simulation with the analytical expression for the SSF as given in the high-density theory.\cite{KM18} As shown in Fig.\ \ref{ssf_theory}, the theoretical SSF calculated from the formula (given in Appendix A for ready reference) and the simulated SSF are in excellent agreement.

\begin{figure}[!h]
\centering
\includegraphics[scale=0.33]{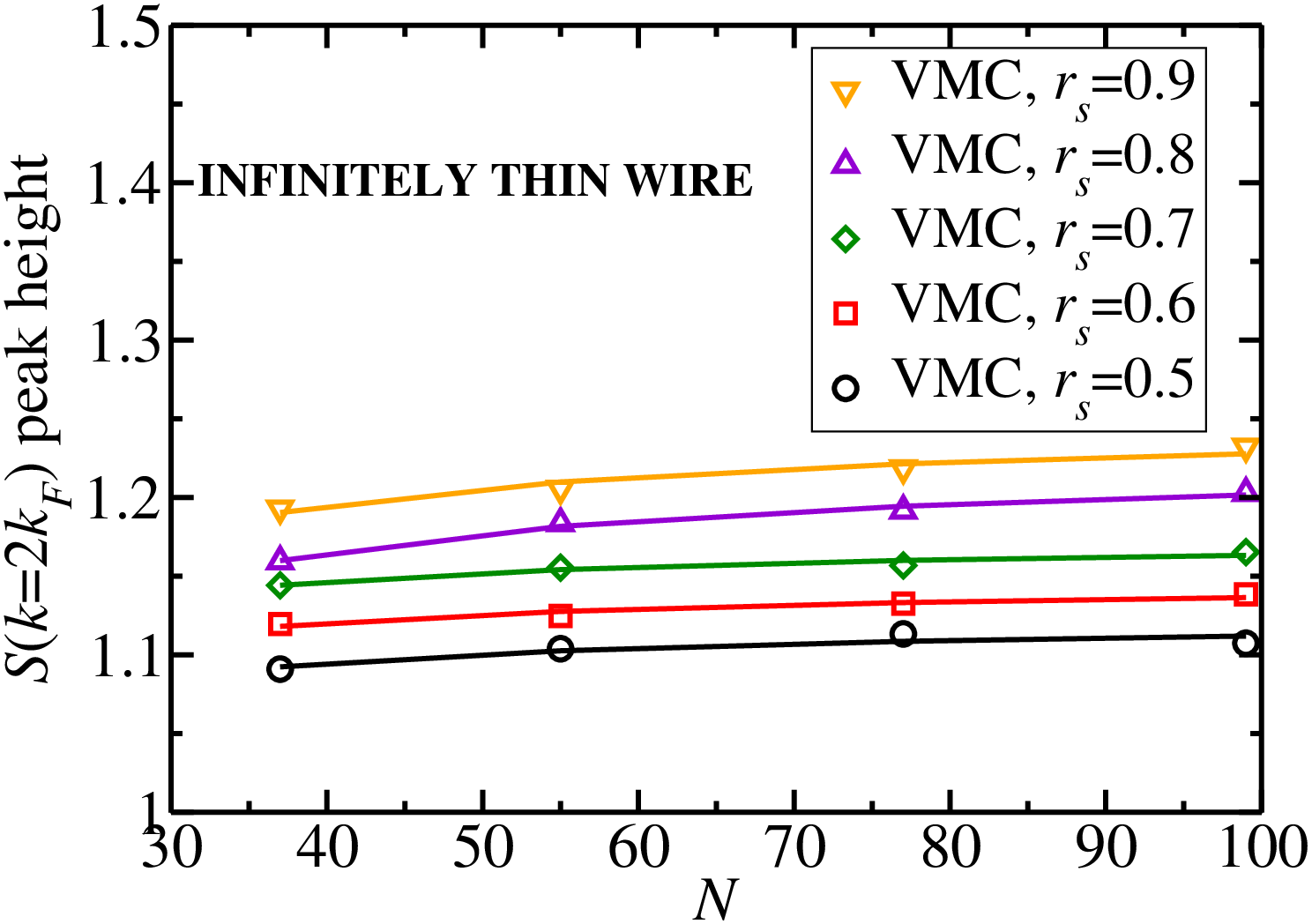}
\caption{(Color online) SSF peak height at $k=2k_F$ plotted against system size $N$ for infinitely thin wires with different coupling parameters $r_s$.}\label{ssf_vs_N}
\end{figure}

\begin{figure}[!h]
\centering
\includegraphics[scale=0.35]{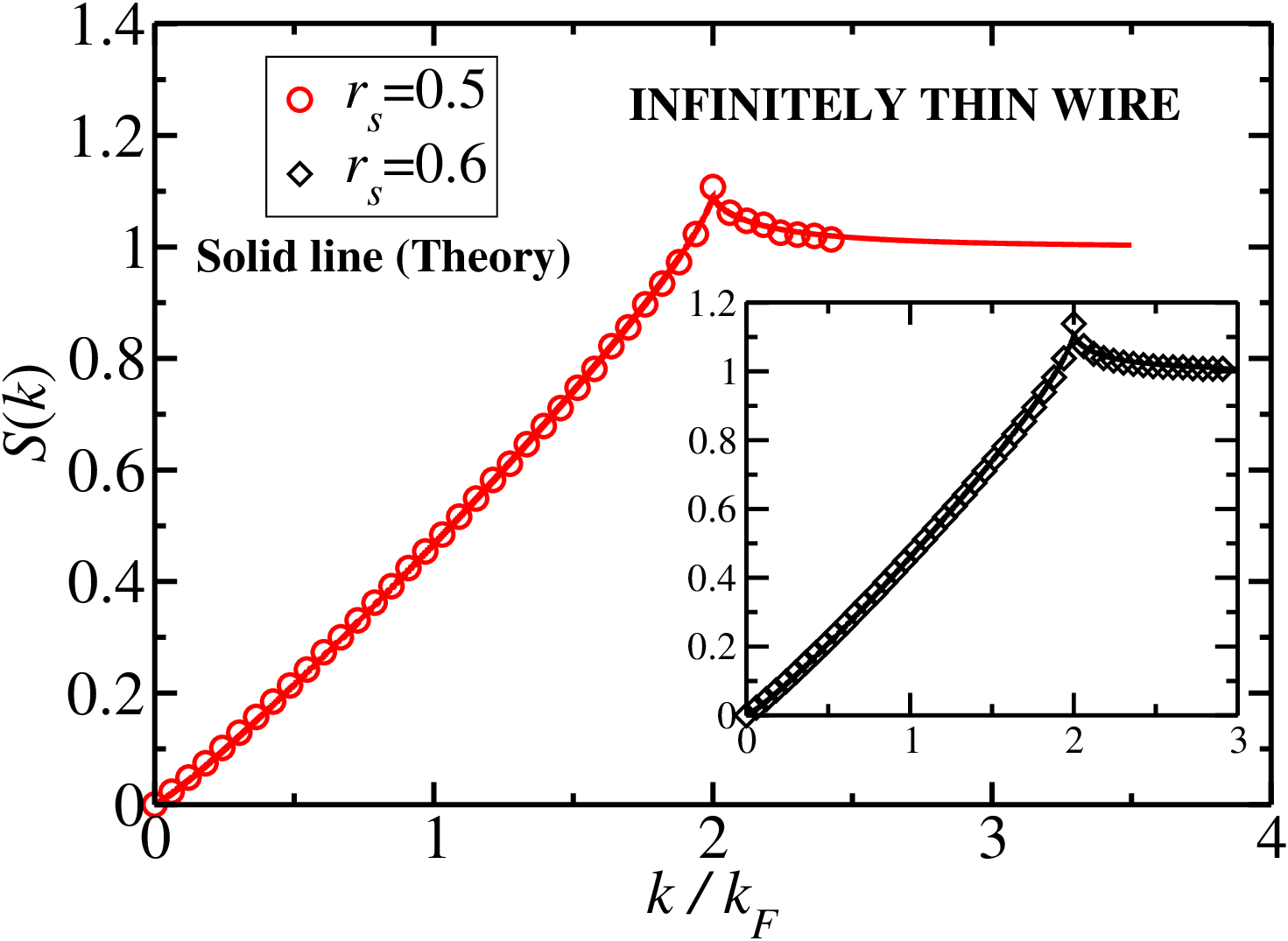}
\caption{(Color online) VMC SSF of an infinitely thin wire with $N=99$, compared with the high-density theory\cite{KM18} (solid line).
The main plot shows the SSF for $r_s=0.5$ and the inset is for $r_s=0.6$.}\label{ssf_theory}
\end{figure}

In Fig.\ \ref{ssf_vs_N} we have plotted the SSF peak height at $k=2k_F$ against $N$, and it is fitted with a function $S(k=2k_F)=a+b/N$.  In the thermodynamic limit the peak height tends to a constant value. For $r_s=0.5$, $0.6$, $0.7$, $0.8$, and $0.9$, the height becomes $a=1.12976$, $1.14735$, $1.17464$, $1.22649$, and $1.25001$, respectively. As we increase the coupling parameter $r_s$, the peak height in the thermodynamic limit is also increased.

{\bf Momentum density:} The MD is a fundamental quantity and it is calculated from the ground-state trial wave function using the formula
\begin{eqnarray}
n(k)=\frac{1}{2\pi}\bigg< \int \frac{\psi_T(r)}{\psi_T(x_1)}\exp[ik(x_1-r)] \, dr\bigg>,
\end{eqnarray}
where the trial wave function, $\psi_T(r)$ is evaluated at $(r,x_2,\ldots,x_N)$. The angular brackets denote the VMC expectation value, obtained as the mean over electron coordinates $(x_1,\dots,x_N)$ distributed as $|\psi_T|^2$.
\begin{figure}[!h]
\centering
\includegraphics[scale=0.31]{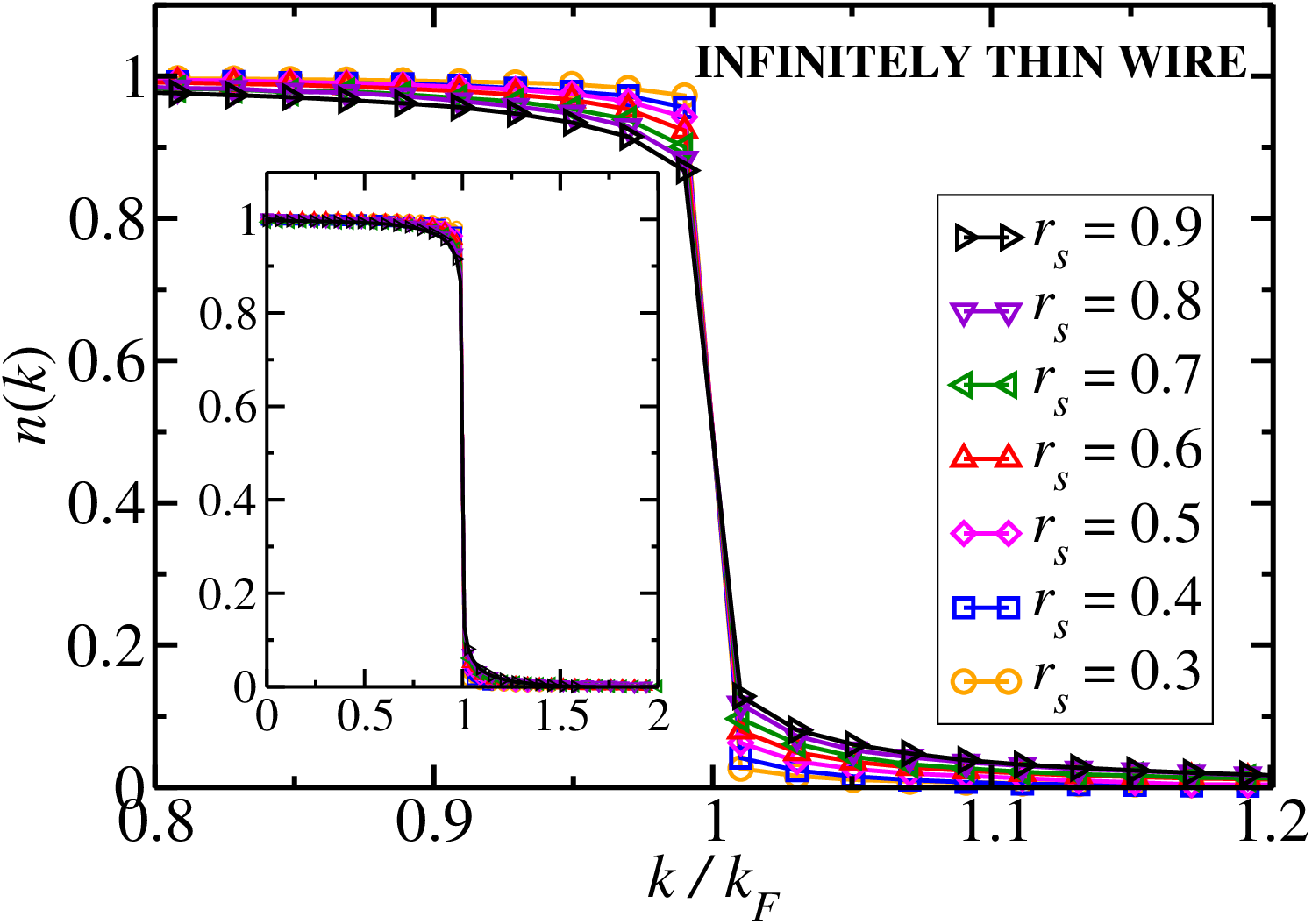}
\vskip 0.4in
\includegraphics[scale=0.31]{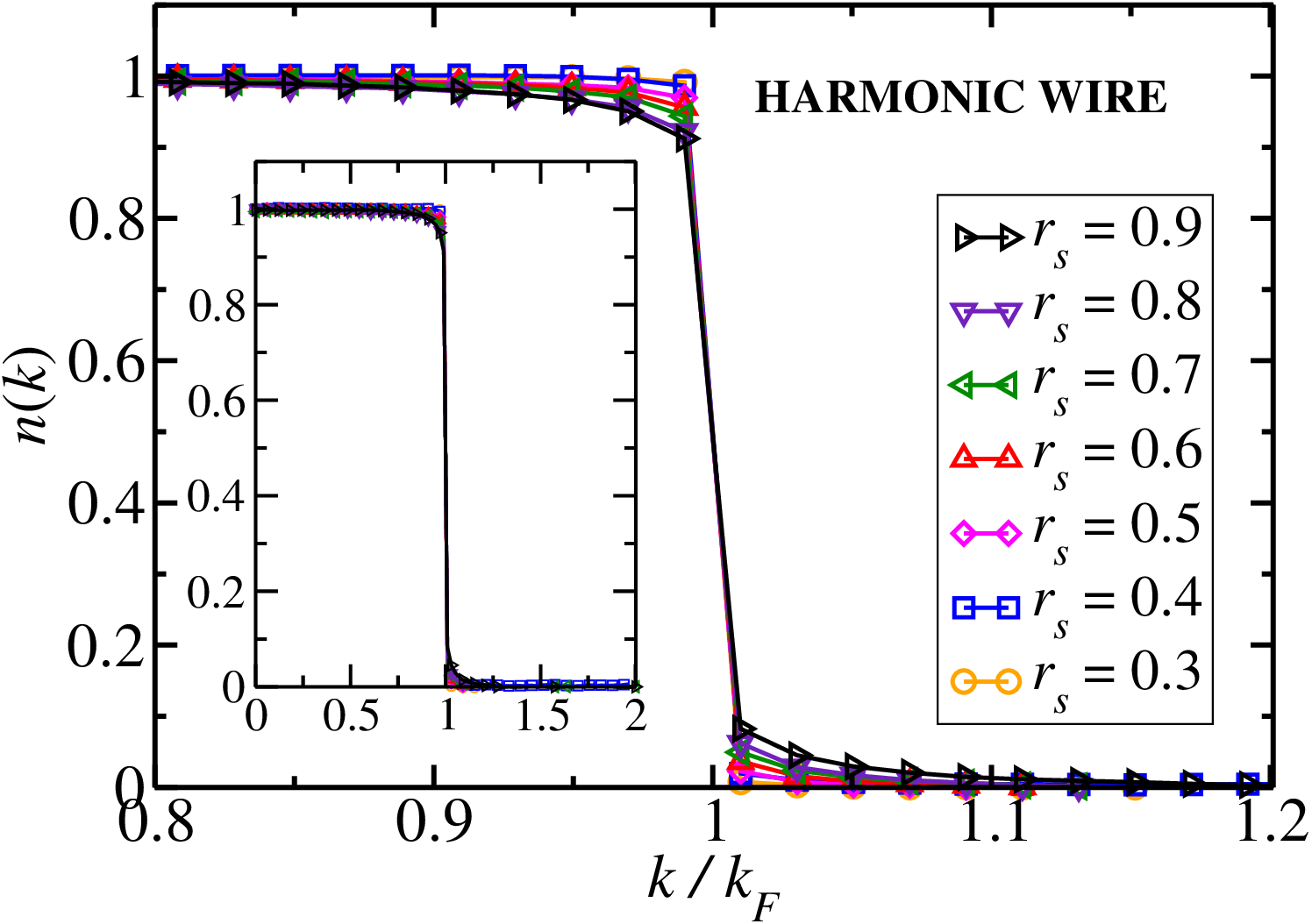}
\caption{(Color online) (upper) MD of an infinitely thin wire and (lower) MD of a harmonic wire of width $b=0.5$ at
several densities. The data shown are for $N = 99$. The statistical error bars are much
smaller than the symbols and have therefore been omitted for
clarity. The inset shows a zoomed-out view.}\label{MD_b0_b0.5}
\end{figure}

In higher dimensions $d>1$, the single-particle states are occupied up to the Fermi energy at $T=0$. So the MD of a state with momentum $k$ has a discontinuity at the Fermi surface. The amplitude of the discontinuity is 1 for a free-electron (noninteracting) system, and the spectral function for free electrons is a delta-function peak. Now when the interaction is switched on, the remarkable Landau Fermi liquid theory applies and the properties of the system remain essentially similar to those of free fermionic particles. The elementary particles are not the individual electrons anymore, but are dressed by the density fluctuations around them and are called quasiparticles. The MD $n(k)$ of a state with momentum $k$ still has a discontinuity at the Fermi wavenumber $k=k_F$ (the spectral function possesses a quasiparticle peak), but with reduced amplitude $Z<1$. On the other hand, in 1D the interaction leads to a power-law behavior in the MD, which is continuous at $k_F$, though the derivative of the MD is singular at $k=k_F$. Near the Fermi wavenumber the TL liquid theory predicts the MD should take the form\cite{Luttinger63,Mattis65}
\begin{equation}
n(k)=n(k_F)+A[{\rm sign}(k-k_F)]|k-k_F|^{\alpha}
\label{nk_MD}
\end{equation}
where $n(k_F)$, $A$, and $\alpha$ are density-dependent parameters. The exponent $\alpha$ may be written in terms of the TL parameter $K_\rho$ as\cite{Schulz90}
\begin{eqnarray}
\alpha=\frac{1}{4}\bigg( K_{\rho}+\frac{1}{K_\rho}-2\bigg). \label{eq:alpha}
\end{eqnarray}

\begin{figure}[!h]
\centering
\includegraphics[scale=0.3]{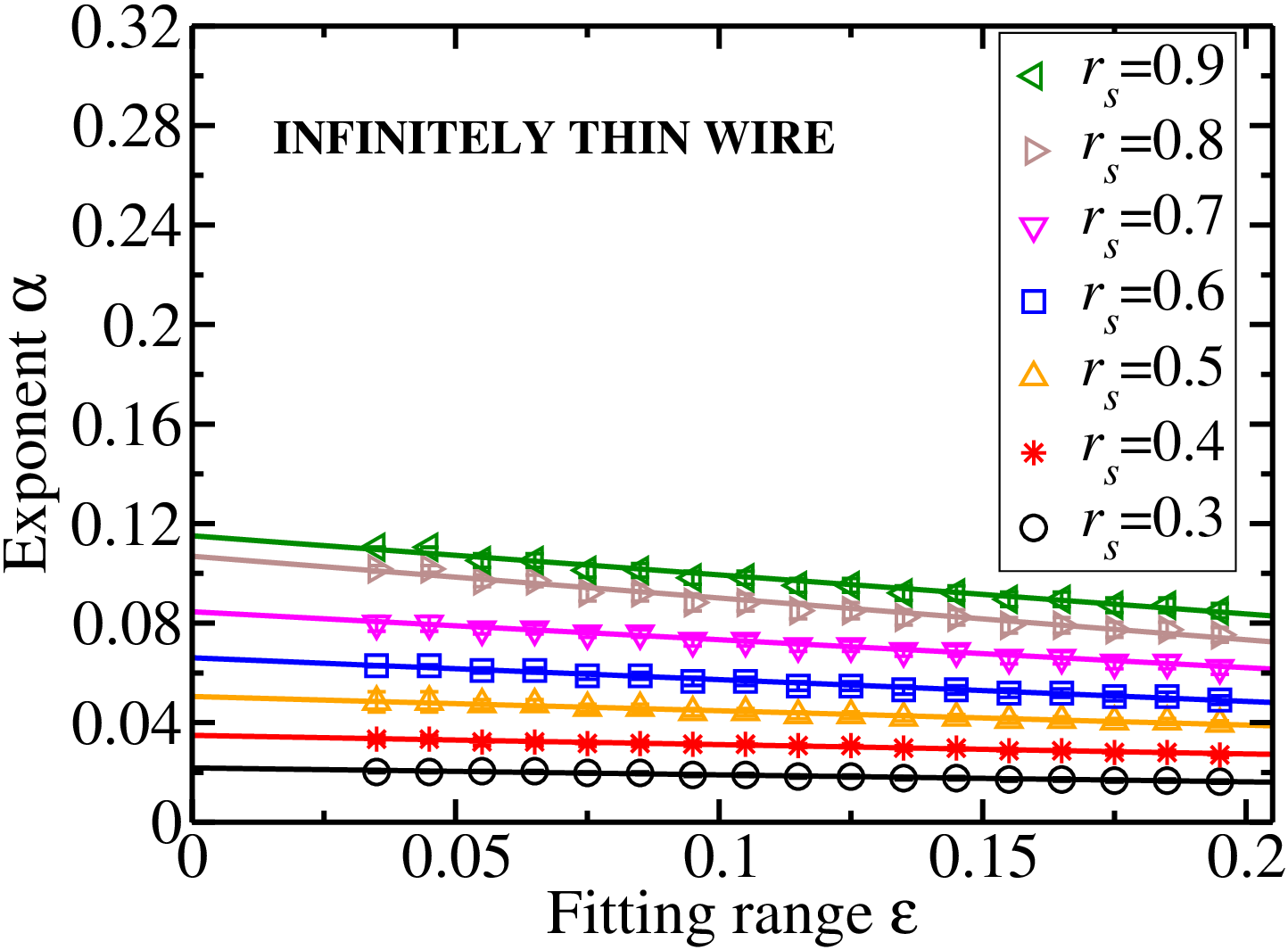}
\vskip 0.3in
\includegraphics[scale=0.3]{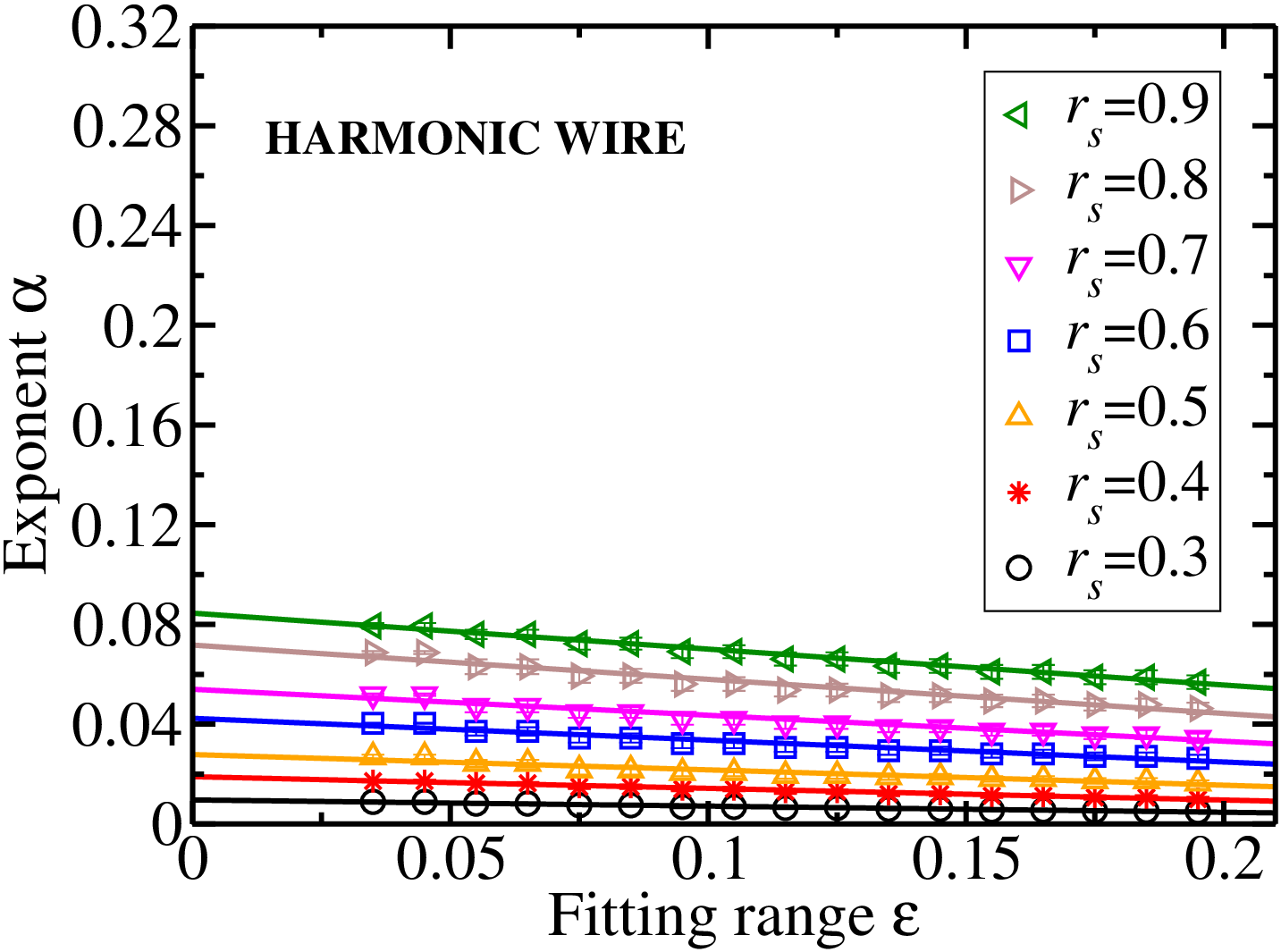}
\caption{(Color online) Tomonaga-Luttinger liquid exponent $\alpha$ in Eq.\ (\ref{nk_MD}) extracted from our MDs against the fitting range of data ($|k-k_F|<\epsilon k_F$), for (upper) a ferromagnetic, infinitely thin wire and (lower) a harmonic wire of width $b=0.5$
The extracted exponent is linearly fitted with the solid line in the region $\epsilon>0.035$ and extrapolated to $\epsilon=0$.}\label{alpha_b0_b0.5}
\end{figure}

\vskip 0.3in

\begin{figure}[!h]
\centering
\includegraphics[scale=0.3]{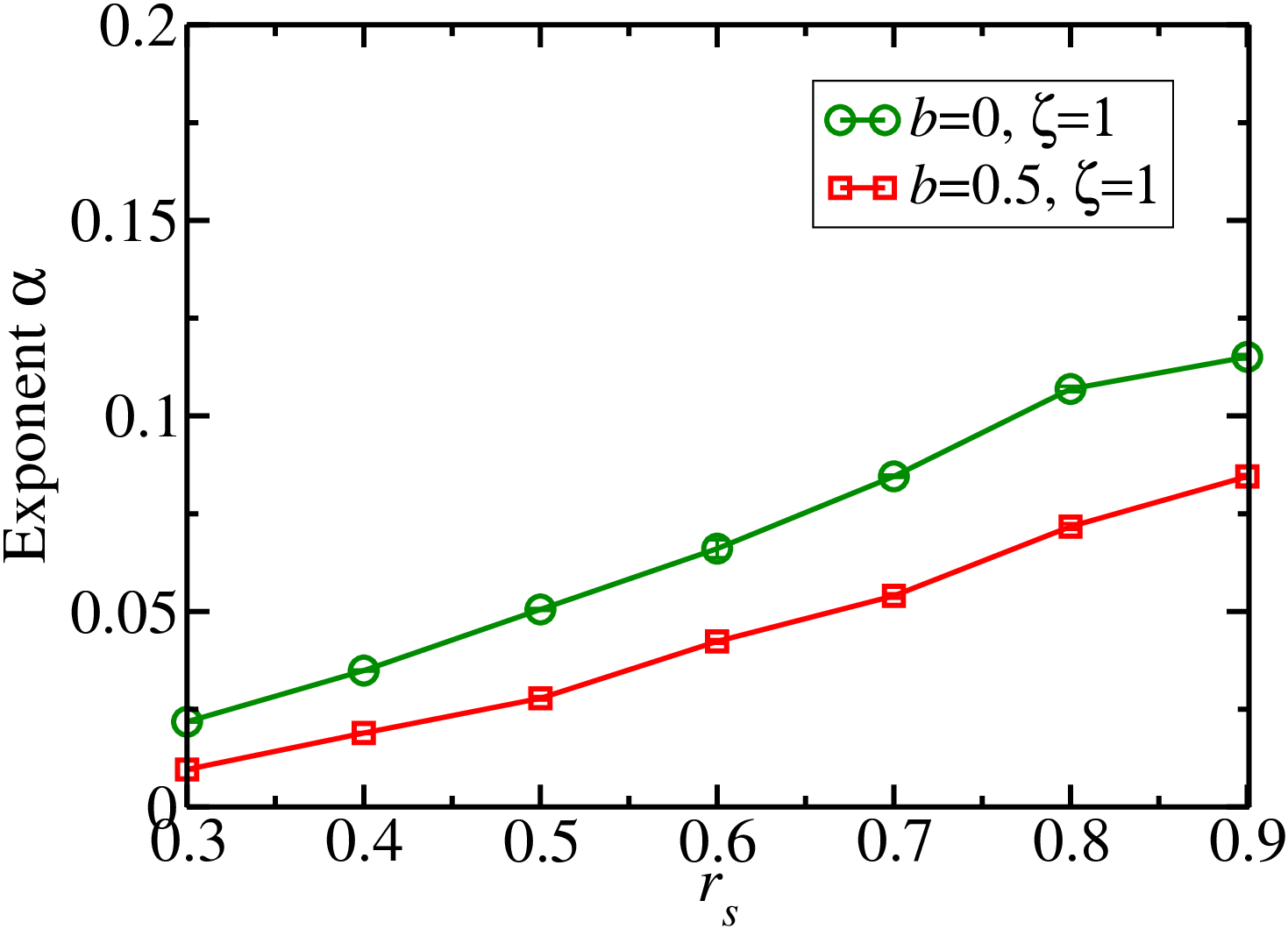}
\caption{(Color online) Exponent $\alpha$, found by fitting Eq.\ (\ref{nk_MD}) to the MDs of ferromagnetic 1D electron gases and extrapolating to $\epsilon=0$, plotted against $r_s$.  The error bars on the data points approximately account for the random error due to the Monte Carlo evaluation of the momentum density of the trial wave function (approximate because the data points in the MD are correlated); however, the random noise on the exponents is clearly larger than these error bars.  There is an additional uncertainty in the MD and hence $\alpha$ due the stochastic optimization of the trial wave function, and this may be responsible for the larger noise.  Nevertheless, the noise in the exponent $\alpha$ as a function of $r_s$ is at least an order of magnitude smaller than the systematic behavior.}\label{alpha_H_Cy_rs}
\end{figure}

\begin{figure}[!hbt]
\centering
\includegraphics[scale=0.32]{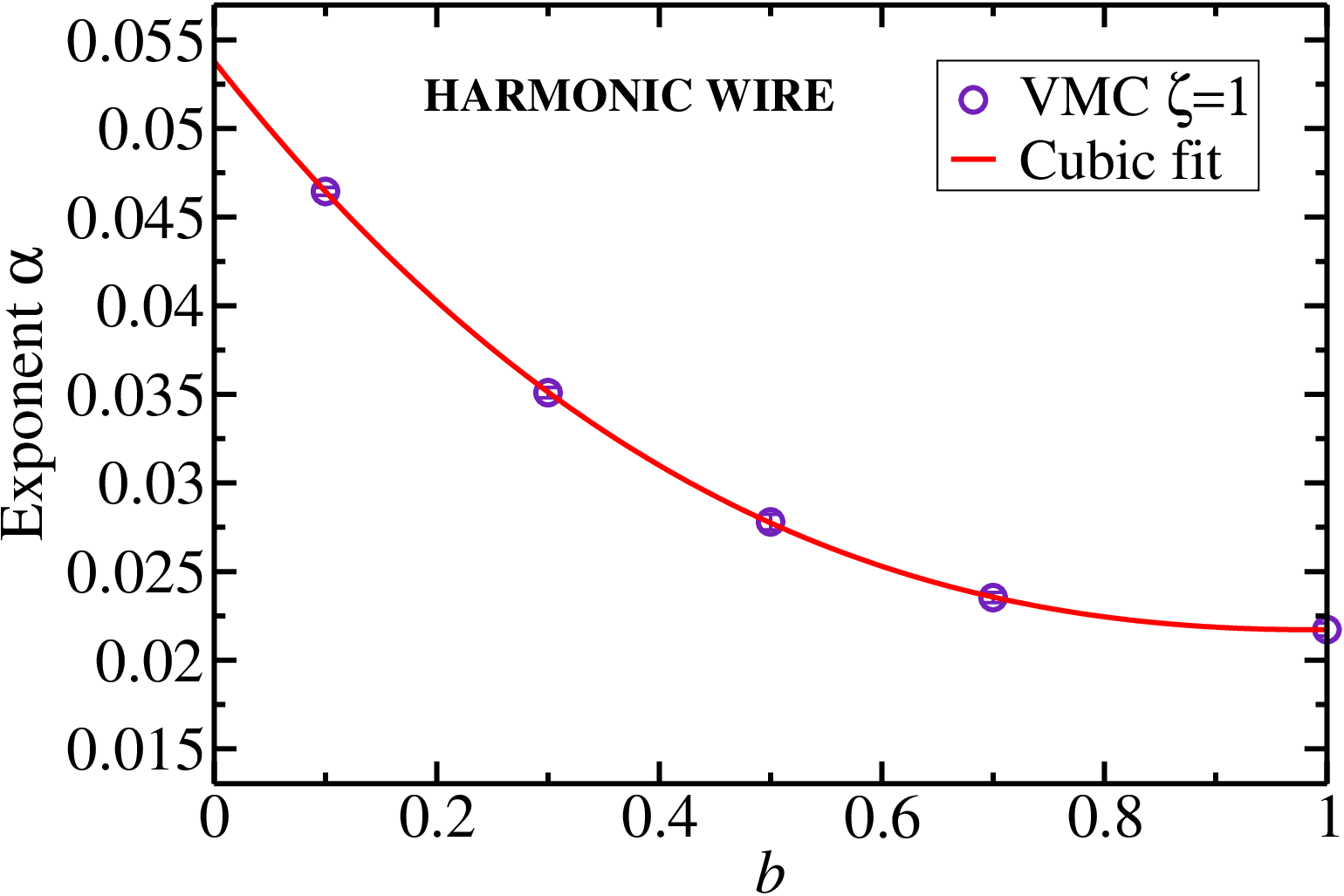}
\caption{(Color online) Exponent $\alpha$ found by fitting Eq.\ (\ref{nk_MD}) to the MDs of ferromagnetic systems for $r_s=0.5$ and different harmonic wire widths $b$.}\label{alpha_H_b}
\end{figure}

\begin{figure}[!h]
\centering
\includegraphics[scale=0.32]{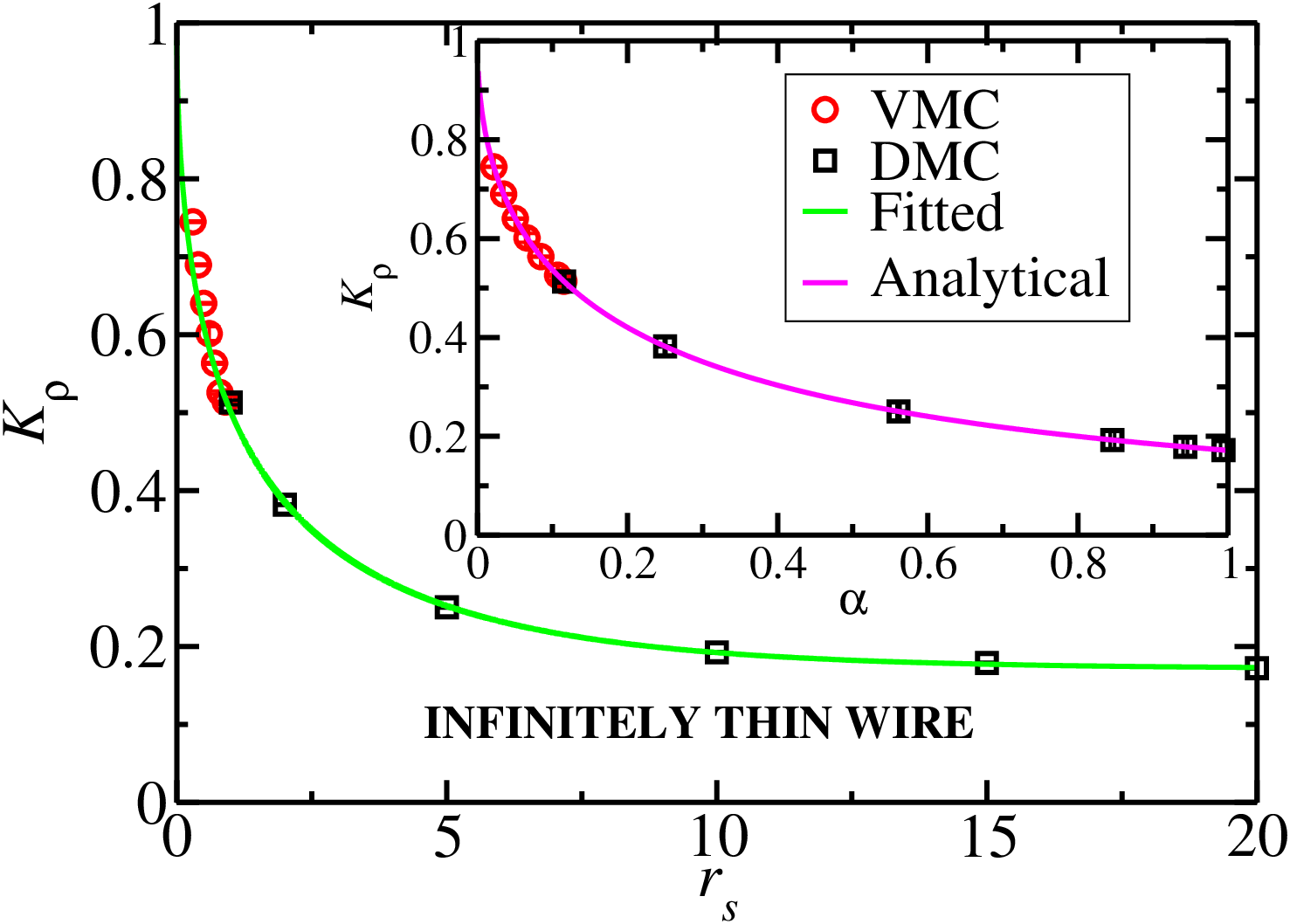}
\caption{(Color online) Tomonaga-Luttinger parameter $K_\rho$ plotted against $r_s$ and, in the inset, plotted as a function of exponent $\alpha$ for an infinitely thin wire. The low density DMC data are adopted from Lee and Drummond.\cite{Lee11}}\label{K_rho_alpha_rs}
\end{figure}

In Fig.\ \ref{MD_b0_b0.5} we have plotted the MD using VMC for an infinitely thin wire ($b=0$) and a harmonic wire of width $b=0.5$.  We have used a single system size, $N=99$, for which the energy is close to the thermodynamic limit; moreover, finite-size effects in the MD appear to be small at low density.\cite{Lee11} To extract the value of the exponent $\alpha$, we have fitted Eq.\ (\ref{nk_MD}) to our MD data near $k_F$. The range of MD data used in the fit is described by $|k-k_F|<\epsilon k_F$. Ideally $\epsilon\rightarrow 0$, as Eq.\ (\ref{nk_MD}) is only valid for $k\rightarrow k_F$. As shown in Fig.\ \ref{alpha_b0_b0.5}, we perform linear extrapolation of $\alpha$ to $\epsilon=0$ by including the points where $\epsilon>0.035$. The trend in the variation of the exponent at high density with respect to $\epsilon$ is similar to what has been observed by Lee and Drummond\cite{Lee11} for lower density. The variation of the exponent $\alpha$ with $r_s$ is plotted in Fig.\ \ref{alpha_H_Cy_rs}. It is found that in the high-density limit $\alpha$ tends to zero, whereas Lee and Drummond\cite{Lee11} have found that in the low-density limit $\alpha$ tends to 1. The exponent $\alpha$ for $r_s=0.5$ as a function of the width $b$ of the wire is shown in Fig.\ \ref{alpha_H_b}. As the width of the harmonic wire decreases the exponent $\alpha$ increases. Using a cubic fit the exponent is extrapolated to $b=0$, giving $\alpha=0.0538(6)$, which agrees with the result for an infinitely thin wire [$\alpha=0.0505(2)$ at $r_s=0.5$].

%

Further, the TL parameter is obtained from Eq.\ (\ref{eq:alpha}) as $K_\rho=1+2\alpha-2\sqrt{\alpha+\alpha^2}$. It is noted that for repulsive interactions $K_\rho$ is positive and $<1$. The value of $\alpha$ is approximately given by $\alpha=\text{tanh}(r_s/8)$.\cite{Lee11} This formula continues to describe the value of $\alpha$ for $r_s<1$ for the infinitely thin wire, although the fractional error increases significantly when $r_s \leq 0.5$.  When substituted in the relation between $K_\rho$ and $\alpha$, this yields
\begin{eqnarray}
K_{\rho} & = & 1+2
   \tanh (r_s/8) \nonumber \\ & & {} -2 \sqrt{\tanh (r_s/8)+\tanh ^2(r_s/8)}.
 \label{Krhors}
\end{eqnarray}
This is plotted in Fig.\ \ref{K_rho_alpha_rs} as function of $r_s$ (and in the inset as a function of $\alpha$).  The low-density data have been taken from Lee and Drummond.\cite{Lee11} The TL parameter $K_\rho$ we obtain for high density smoothly goes over to the value we obtained for low density from the data of Lee and Drummond.\cite{Lee11} The TL parameter $K_\rho$ is well represented by the formula given in Eq.\ (\ref{Krhors}) for the infinitely thin wire.

In the present work the TL liquid behavior is characterized by the power-law decay in the momentum distribution function at the Fermi wavenumber. The TL parameter $K_\rho$ gives a quantitative value of the correlation strength. Small values of the TL parameter $K_\rho$ imply a strongly correlated system. The results in Fig.\ \ref{K_rho_alpha_rs} immediately indicate non-Fermi liquid behavior. In other words, we confirm in the present study that howsoever small the interaction may be, the electron fluid in 1D behaves as a TL Liquid. We further observed that the structure factor at $k=2k_F$ has a peak even in the high-density regime.

Our high-density simulation predicts an exponent $\alpha$ between 0.02 and 0.12, whereas in the low-density regime\cite{Lee11} it ranges from 0.12 to $\sim 1$. There have been experiments in the low-density regime, e.g., in carbon nanotubes, but experimental results for the high-density 1D electron fluid are not yet available.


\section{Conclusions}
\label{Conclusion}

We have performed a detailed VMC study of 1D electron fluids interacting via long-range Coulomb potentials for infinitely thin and harmonic wires at high density. For the infinitely thin and harmonic-wire models, we have reported the VMC ground-state energy in the thermodynamic limit. Using the VMC ground-state energy we have calculated the correlation energy. The predicted correlation energy is in agreement with conventional perturbation theory results\cite{Loos13} at high densities. The calculated SSF shows a peak at $2k_F$. The VMC SSF and PCF data show very good agreement with the high-density theory.\cite{KM18}


For $r_s<1$ we have reported VMC results for the MD as function of wavenumber $k$, and the data are used to predict the Tomonaga-Luttinger parameter for infinitely thin and harmonic wires. It is found that the exponent $\alpha$ ranges from 0.02 to 0.12 in the high-density limit. The exponent has been used to obtain the Tomonaga-Luttinger parameter $K_\rho$ as a function of $r_s$. It is hoped that our work will motivate experimental work on high-density 1D electron systems. One example of a suitable material for study is a zigzag carbon nanotube placed on a SrTiO$_3$ substrate.


\begin{acknowledgments}
The authors (V.A.\ and K.N.P.)\ acknowledge the financial support
by National Academy of Sciences of India, Allahabad. The high performance computing
centre facilities at Panjab University has been used to run the
\textsc{casino} code for our QMC calculations.
\end{acknowledgments}

\appendix*
\section{A}
The analytical expression for the SSF in the high-density theory\cite{KM18}
 for $x<1$ is
\begin{eqnarray}
S(k)&=&\frac{x}{2}+\frac{g_s^2 r_s }{\pi ^2 x}\bigg\{(x-2) \ln \left(\frac{2-x}{2}\right) [\ln (4-2 x)\nonumber\\
& &-2 (\ln (x)+1)]+(x+2) \ln
   \left(\frac{x+2}{2}\right)\nonumber\\
   & &\times [2 \ln (x)-\ln (2 x+4)+2]\bigg\},
\end{eqnarray}
where $x=k/k_F$, and for $x>1$
\begin{eqnarray}
S(k)&=&1+\frac{g_s^2 r_s}{\pi ^2 x} \bigg\{(2-x) \ln ^2(x-2)-(x+2) \ln ^2(x+2)\nonumber\\
& &+2 (x-2) [\ln (x)+1] \ln (x-2)-2 x \ln (x) [\ln (x)+2]\nonumber\\
& &+2 (x+2)
   [\ln (x)+1] \ln (x+2)\bigg\}.
\end{eqnarray}
The PCF $g(r)$ is obtained from the SSF $S(k)$ as
\begin{eqnarray}
g(r)=1-\frac{1}{2\pi n}\int^{\infty}_{-\infty} dk\; e^{ikr}[1-S(k)].
\end{eqnarray}

\end{document}